\documentclass[preprint,showpacs,preprintnumbers]{revtex4}
\usepackage{bm,enumerate,dcolumn,tikz,graphicx,color,amsmath,amssymb}
\usepackage{epstopdf}
\usepackage{empheq,xfrac}
\usepackage{capt-of}
\usepackage{pgfplots}

\usepgfplotslibrary{colormaps,external}
\usetikzlibrary{decorations.markings}
\usetikzlibrary{arrows}

\newcommand{\comment}[1]{}
\newcommand{\bra}{\langle}
\newcommand{\ket}{\rangle}

\newcommand{\velg}{{\textsc{\tiny VG}}}

\newcommand{\sage}{http://etheses.dur.ac.uk/849/}

\newcommand{\Rmnum}[1]{\expandafter\@ slowromancap\romannumeral #1@}
\newcommand{\RNum}[1]{\uppercase\expandafter{\romannumeral #1\relax}}
\definecolor{lightblue}{rgb}{.8, .8, 1}
\definecolor{myyellow}{RGB}{254,241,24}
\definecolor{myorange}{RGB}{234,125,1}
 
\usepackage{array}
\newcolumntype{C}[1]{>{\centering\let\newline\\\arraybackslash\hspace{0pt}}m{#1}}
\newcolumntype{L}[1]{>{\raggedright\let\newline\\\arraybackslash\hspace{0pt}}m{#1}}
\newcolumntype{R}[1]{>{\raggedleft\let\newline\\\arraybackslash\hspace{0pt}}m{#1}}

\newcommand{\threej}[6]{
\begin{pmatrix}
  #1 & #2 & #3\\
  #4 & #5 & #6
\end{pmatrix}
}
\newcommand{\sixj}[6]{
\begin{Bmatrix}
  #1 & #2 & #3\\
  #4 & #5 & #6
\end{Bmatrix}
}

\begin{document}

\title{Divalent Rydberg atoms in optical lattices: intensity landscape and magic trapping} 
\author{Turker Topcu and Andrei Derevianko}
\affiliation{Department of Physics, University of Nevada, Reno, NV 89557, USA}
\date{\today}

\begin{abstract}

We develop a theoretical understanding of trapping divalent Rydberg atoms in optical lattices. Because the size of the Rydberg electron cloud can be comparable to the scale of spatial variations of laser intensity, we pay special attention to averaging optical fields over the atomic wavefunctions. Optical potential is proportional to the ac Stark polarizability. We find that in the independent particle approximation for the valence electrons, this polarizability breaks into two contributions: the singly ionized core polarizability and the contribution from the Rydberg electron. Unlike the usually employed free electron polarizability, the Rydberg contribution depends both on laser intensity profile and the rotational symmetry of the total electronic wavefunction. We focus on the $J=0$ Rydberg states of Sr and evaluate the dynamic polarizabilities of the 5s$n$s($^1S_0$) and 5s$n$p($^3P_0$) Rydberg states. We specifically choose Sr atom for its optical lattice clock applications. We find that there are several magic wavelengths in the infrared region of the spectrum at which the differential Stark shift between the clock states (5s$^2$($^1S_0$) and 5s5p($^3P_0$)) and the $J=0$ Rydberg states, 5s$n$s($^1S_0$) and 5s$n$p($^3P_0$), vanishes. We tabulate these wavelengths as a function of the principal quantum number $n$ of the Rydberg electron. We find that because the contribution to the total polarizability from the Rydberg electron vanishes at short wavelengths, magic wavelengths below $\sim$1000 nm are ``universal" as they do not depend on the principal quantum number $n$. 

\end{abstract}


\maketitle

\section{Introduction}

Quantum information processing (QIP) with neutral atoms has a number of distinct and appealing advantages, such as scalability, massive parallelism, long coherence times and reliance on well-established experimental techniques. Historically, the dominant fraction of QIP schemes with neutral atoms has focused on alkali-metal atoms, which possess a single valence electron outside a tightly-bound atomic core. The most successful experimental demonstration~\cite{UrbJohHen09,GaeMirWil09}  of quantum two-qubit logic gate with neutral atoms has been carried out using Rydberg gates, originally proposed in Ref.~\cite{JakCirZol00}. These experiments employed Rydberg excitations of $^{87}$Rb, an alkali-metal atom. In recent years, there have been important new developments with cooling and trapping divalent atoms, such as group-II atoms ({\it e.g.}, Mg, Ca, Sr) and group-II-like atoms such as Yb, Hg, Cd, and Zn, which can greatly benefit experiments. Considering the experimental success of Rydberg gates with alkali-metal atoms, it is natural to ask if the distinct properties of divalent atoms could improve the experimental feasibility~\cite{GilMukBri13,LocBodSad13,ValJonPot12,MukMilNat11,MilLocCor11}. 

Optical trapping is essential for QIP experiments due to long coherence times that can be achieved. In QIP experiments, the size of Rydberg atoms can easily be larger than the lattice constant of the optical lattice. As we recently demonstrated in~\cite{TopDer13,TopDerTuneOut13}, one-dimensional trapping potential for an alkali-metal Rydberg atom is proportional to the expectation value $\bra\cos(2kz)\ket$, where $k=2\pi/\lambda$ is the wavenumber of the optical lattice laser whose wavevector is aligned with $z$-axis. Here $\lambda=2\pi c /\omega$ is the lattice laser wavelength, $\omega$ is frequency and $c$ is the speed of light. We have termed $\alpha^{\rm lsc}_{r}(\omega)=-\bra\cos(2kz)\ket/\omega^2$ ``landscaping polarizability" as it modulates the free-electron polarizability $\alpha_{\rm e}(\omega)=-1/\omega^2$ according to the intensity profile of the optical lattice.  The factor $\bra\cos(2kz)\ket$ has a universal dependence on $n^2 a_0/\lambda$ and in the short wavelength $\lambda$ (or high principal quantum number $n$) limit, $n^2 a_0/\lambda \gg 1$ and $\alpha^{\rm lsc}_{r}(\omega)\rightarrow 0$. In this limit, the Rydberg atom is no longer trapped as the optical trapping potential is directly proportional to the landscaping polarizability. Divalent Rydberg atoms can still remain trapped even in this limit. 

Indeed, the divalent atoms have the advantage of a second optically active valence electron, which contributes to the total polarizability, making it easier to trap the atoms in a tight optical lattice. Optical trapping of Rydberg atoms also faces challenges associated with the small polarizability of Rydberg  states. In this sense, utilization of divalent atoms can greatly simplify the trapping of Rydberg states because the second (non-Rydberg) valence electron sizably contributes to the trapping potential, which is proportional to the atomic polarizability. Due to the resonant structure of polarizability contributed by the non-Rydberg valence electron, the lattice wavelength can be also tuned so as to make the total trapping potential large~\cite{MukMilNat11}. 
 
QIP with neutral trapped atoms comes at a price: trapping optical fields strongly perturbs atomic energy levels such that there are uncontrollable accumulations of differential phase between qubit states as atoms move in the traps~\cite{SafWal03,SafWal10}. In addition, the underlying Stark shift is proportional to the local intensity of the trapping lasers; the shift is non-uniform across the atomic ensemble and it is also sensitive to laser intensity fluctuations. This problem is elegantly mitigated using the so-called ``magic" traps~\cite{DerKat11}. At the ``magic" trapping conditions, two atomic levels of interest are shifted by exactly same amount by the trapping fields; therefore the differential effect of trapping fields simply vanishes for that qubit transition. The idea of such “magic” trapping has also been crucial for establishing a new class of atomic clocks: the optical lattice clocks~\cite{DerKat11}. 

When we go back to neutral alkali-metal atoms and review the literature, we easily find a large body of work on optical trapping and utilization of Rydberg states in quantum information experiments~\cite{BreCavJes99,JakBriCir99,YouCha2000,LukHem2000,MomEckErt03}. However, the situation is different for divalent atoms and the ideas on involving them in experiments are in their infancy. To our knowledge, two papers so far have considered divalent systems in this setting. Mukherjee {\it et al.} in Ref.~\cite{MukMilNat11} considered the possibility of studying many-body physics using alkaline-earth atoms trapped in optical lattices. They found magic wavelengths for simultaneously trapping $5sns$($^1S_0$) Rydberg states with the $5s^2$ ground state for the Sr atom. In their treatment, the polarizability in the Rydberg state is evaluated by adding the polarizability of the Sr$^+$ ion to the free electron polarizability, representing the contribution from the Rydberg electron. This treatment neglects averaging of laser intensity profile over the Rydberg wavefunction. Also in Ref.~\cite{MukMilNat11}, the splitting into core (Sr$^+$) and Rydberg polarizabilities has been carried out in an {\it ad hoc} manner. Our rigorous derivation presents below shows that indeed for the $5sns$($^1S_0$) states, the total polarizability splits into independent contributions from the valence and the Rydberg electrons. We will show in this paper that the situation is more complicated in the more general case, and even in the independent-particle approximation, the extent to which the Rydberg electron can contribute to the total polarizability is dictated by the angular momentum $J$ of the entire atom. Furthermore, in a recent paper~\cite{TopDer13}, we demonstrated that the free-electron polarizability of the Rydberg electron is modulated by the intensity distribution in an optical lattice, and it is this ``intensity landscape modulated polarizability" that plays a role in trapping the Rydberg state. It can have both positive and negative values and, in contrast to the treatment of Ref.~\cite{MukMilNat11}, is not simply the free-electron polarizability. In another paper~\cite{OvsDerGib11}, Ovsiannikov {\it et al.} proposed using optically trapped Sr atoms in Rydberg states to probe ambient temperature at 10 mK level in clock experiments. To this end, they showed that the ac Stark shift experienced by the Rydberg electron is modulated by the intensity distribution. In their treatment, the contribution from the Sr$^+$ ionic core was neglected. Here we combine the complimentary treatments of Refs.~\cite{MukMilNat11} and~\cite{OvsDerGib11} and rigorously derive and evaluate ac polarizabilities of divalent Rydberg atoms. 

In alkali-metal atoms, one can always find magic wavelengths above a certain $n$~\cite{TopDer13} and we find that the same holds true for divalent atoms. We find and tabulate several wavelengths at which magic trapping conditions can be attained for the clock states $5s^2$($^1S_0$) and $5s5p$($^3P_0$) of Sr with $J=0$ Rydberg states. 

The paper is organized as follows: in Sec.~\ref{sec:magic_lands} we start by briefly reviewing optical trapping of  alkali-metal Rydberg atoms and develop a theoretical understanding using the second order perturbation theory for divalent Rydberg atoms. We break down atomic polarizability individual contributions from the valence and the Rydberg electrons taking the rotational symmetry of the many-body state into account. In Sec.~\ref{subsec:ion_rydberg}, we evaluate the landscaping polarizability of the Rydberg electron in an $ns$ state and contrast it with the 5s ground state polarizability of the Sr$^+$ ion. In order to find magic wavelengths for $5s^2(^1S_0) - 5sns(^1S_0)$, $5s^2(^1S_0) - 5snp(^3P_0)$, $5s5p(^3P_0) - 5sns(^1S_0)$ and $5s5p(^3P_0) - 5snp(^3P_0)$ transitions, we need to accurately calculate the 5s$^2$ and 5s5p($^3P_0$) state polarizabilities. We perform these calculations in Sec.~\ref{subsec:clock} and show that  we can recover the well known magic wavelength at 814 nm at which Sr optical lattice clocks are operated. We then discuss and calculate the Rydberg state polarizabilities for the 5s$n$s($^1S_0$) and 5s$n$p($^3P_0$) states using that for the Sr$^+$ ion and the contributions from the Rydberg electron individually in Sec.~\ref{subsec:divalent_states}. Finally, we demonstrate that the magic trapping conditions for these Rydberg states with the clock states can be satisfied at several wavelengths in the IR range and below $\sim 1400$ nm. Universal ($n$-independent) magic trapping is shown to exist for $\lambda<1000$ nm. We conclude in Sec.~\ref{sec:conclusions} with final remarks. Unless specified otherwise, atomic units, $|e|=\hbar=|m_e|\equiv 1$ are used throughout the paper. We also use the Gaussian system of units for electromagnetic quantities.

\section{Formalism}\label{sec:magic_lands}
In this section, we develop a formalism for computing adiabatic trapping potentials for divalent atoms. The formalism requires both understanding of how Rydberg electron wavefunction averages over lattice laser intensity (landscaping polarizability) and the many-body character of the two-electron states of specific rotational symmetry. In Sec.~\ref{subsec:magic_alkali}, we start by reviewing main ideas behind landscaping polarizability for alkali-metal Rydberg atoms introduced in Refs.~\cite{TopDer13,TopDerTuneOut13}. In Sec.~\ref{subsec:magic_divalent}, we discuss atomic structure for two-electron states and then move onto deriving a second order perturbative expression for the  polarizability of divalent Rydberg atoms in the velocity gauge. We particularly pick Sr atom due to its wide spread use in optical lattice clocks and well developed experimental techniques for its cooling and trapping in the ground state. 

\subsection{Optical trapping and alkali metal atoms}\label{subsec:magic_alkali}
In a recent paper~\cite{TopDer13}, we demonstrated that the trapping potential $U_r(Z)$ seen by alkali-metal Rydberg atoms in an optical lattice formed along the $z$-axis can be decomposed into a position dependent term which varies as the position of the atom $Z$ changes along the optical lattice, and an offset term $U^0_r$, which shifts the potential energy by a fixed amount everywhere along the lattice: 
\begin{equation}\label{eq:pot_split}
U_r(Z) = U^0_r + U^Z_r \sin^2(kZ) \;. 
\end{equation} 
Here $k=\omega/c=2\pi/\lambda$ is the lattice laser wave vector. It is the $Z$-dependent piece of the potential $U_r(Z)$ that provides confinement in the $z$-direction, because only this part of the potential can exert force on the atom. For a Rydberg state $|r\ket=|nlm_z\ket$, the position  dependent term $U^Z_r$ and the offset $U^0_r$ in the trapping potential can be written as 
\begin{eqnarray}\label{eq:pot_zdep}
U^Z_r &=& \frac{F_0^2}{4\omega^2} \bra n l m_z|\cos (2k z)|n l m_z \ket 
	\equiv- \alpha^{\rm lsc}_{nlm_z}(\omega) \frac{F_0^2}{4} \; , \\
U^0_r &=& \frac{F_0^2}{4\omega^2} \bra n l m_z|\sin^2 (k z)|n l m_z \ket \; , 
\label{eq:pot_offset}
\end{eqnarray}
where $z$ is the position of the Rydberg electron relative to the nucleus and $F_0$ is the laser field strength. We termed $\alpha^{\rm lsc}_{nlm_z}(\omega)$ the landscaping polarizability because it convolutes the free electron polarizability $\alpha_{\rm e}(\omega)=-1/\omega^2$ with the laser intensity profile: 
\begin{equation}
\alpha^{\rm lsc}_{r}(\omega)=\bra\cos(2kz)\ket \alpha_{\rm e}(\omega) \;. 
\end{equation}
In the limit $\bra r\ket \sim a_0 n^2 \ll\lambda$, $U^0_r\rightarrow 0$ and $U^Z_r\rightarrow -\alpha_{\rm e}(\omega)F_0^2/4$. Away from this limit, however, $\alpha^{\rm lsc}_{nlm_z}(\omega)$ exhibits oscillatory behavior and changes sign several times as $\lambda$ is increased before eventually the free electron character takes over. 

The landscaping polarizability can be evaluated by expanding $\cos(2kz)$ in terms of irreducible tensor operators (ITO) $t^{(K)}_{M_{K}}$ of rank $K$: 
\begin{align} \label{eq:alkali_tk} 
\cos(2kz) &= \sum_{K={\rm even}} t^{(K)}_{M_K=0}(r) \;, \\ \label{eq:alkali_tk_def}
t^{(K)}_{M_K}(r) &=  (-1)^{K/2} (2K+1) j_K(2kr) C^{(K)}_{M_K}(\hat{r}) \;. 
\end{align} 
Here $C^{(K)}_{M_K}(\hat{r})=\sqrt{4\pi/(2K+1)}\;Y^{(K)}_{M_K}(\hat{r})$ are the normalized spherical harmonics and $j_K(2kr)$ are the spherical Bessel functions. Application of the Wigner-Eckart theorem results in 
\begin{eqnarray}\label{eq:alkali_alp}
\alpha^{\rm lsc}_{nlm_z}(\omega) &\equiv& -\frac{1}{\omega^2} 
		\bra n l m_z| \sum_{K={\rm even}} t^{(K)}_{M_K=0} |n l m_z \ket \\ 
			&=& -\frac{1}{\omega^2} \sum_{K={\rm even}} (-1)^{l-m_z} \threej{l}{K}{l}{-m_z}{0}{m_z} 
					\bra n l ||t^{(K)} ||n l \ket \;, 
\end{eqnarray}
where the reduced matrix element is given by 
\begin{eqnarray}\label{eq:alkali_rmatel}
\bra n l ||t^{(K)} ||n l \ket 
	= \bra n l ||C_{K} ||n l \ket \int_0^\infty dr P_{nl}^2(r) j_{K}(2kr) \;. 
\end{eqnarray}
Here $P_{nl}(r)$ are radial orbitals and the reduced matrix elements $\bra n l ||C_{K} ||n l \ket$ can be expressed in terms of the 3-$j$ symbols~\cite{VarMosKhe89}. We have shown in~\cite{TopDer13} that as a function of $\omega$, the landscaping polarizability $\alpha^{\rm lsc}_{nlm_z}(\omega)$ for Rydberg states exhibits sizable oscillations in the infrared (IR) region, changing sign several times before the free-electron character $-1/\omega^2$ starts to dominate it. This enables magic trapping conditions for the Rydberg and the ground states of alkali-metal atoms, where $\alpha^{\rm lsc}_{nlm_z}(\omega)$ and the ground state polarizabilities match. Since $\alpha^{\rm lsc}_{nlm_z}(\omega)$ has to vanish in order to change sign, there are also wavelengths at which $\alpha^{\rm lsc}_{nlm_z}(\omega)$ vanishes, which are referred to as the tune-out wavelength~\cite{TopDerTuneOut13}. The Rydberg atom does not feel the optical trap in lattices tuned to these ``tune-out" wavelengths. 

\subsection{Divalent atoms}\label{subsec:magic_divalent}
As we move onto multi-valent Rydberg atoms, a new effect appears. The non-Rydberg (spectator) valence electron can sizably contribute to the total polarizability of the atom. This contribution to the total polarizability of divalent atoms has been taken into account in an {\it ad hoc} manner in Ref.~\cite{MukMilNat11}. Also, Ref.~\cite{MukMilNat11} used the free electron polarizability $\alpha_{\rm e}(\omega)$ to represent the contribution from the Rydberg electron. In this section, we employ the more rigorous concept of landscaping polarizability to evaluate the contribution from the Rydberg electron. Moreover, we develop a theoretical framework which accounts for the overall rotational symmetry of the wavefunction when the atom is in a given $J$-state. It turns out that for states with a Rydberg $s$-electron ({\it e.g.} $5sns(^1S_0)$ for Sr), individual contributions from the ground ($5s$) and the Rydberg ($ns$) states simply add to give the total polarizability of the atom. However, for states with a $p$-electron in the Rydberg state ({\it e.g.} $5snp(^3P_0)$ for Sr), the total rotational symmetry ($J=0$) imposes some restrictions. 

\subsubsection{Atomic structure}\label{subsubsec:magic_structure}
We begin with the two-electron wave function: for a particular rotational symmetry the wavefunction can be expanded in terms of two-particle basis functions as 
\begin{equation} \label{eq:divalent_wf}
\Psi(\pi JM_J) = \sum_{k \ge l} c_{kl} \Phi_{kl}(\pi JM_J) \;.
\end{equation}
Here $J$ is the total angular momentum with projection $M_J$, and $\pi$ is the parity of the state $\Psi$. The basis functions are defined in the subspace of virtual orbitals 
\begin{equation} \label{eq:divalent_wf_exp}
\Phi_{kl}(\pi JM_J) = \eta_{kl} \sum_{m_k,m_l} C^{J M_J}_{j_k m_k j_l m_l} 
	a^{\dagger}_{n_k j_k m_k} a^{\dagger}_{n_l j_l m_l} |0_{\rm core}\ket \equiv |kl(J,M_J)\ket \;,
\end{equation}
where $\eta_{kl}^2 = 1 - \frac{1}{2}\delta_{n_k n_l} \delta_{j_k j_l}$ is a normalization factor, $a^{\dagger}$ are creation operators, and the quasi-vacuum state $|0_{\rm core}\ket$ corresponds to the closed-shell core. The Clebsch-Gordan coefficients $C^{J M_J}_{j_k m_k j_l m_l}$ mix single-particle orbitals to form a wavefunction with a well defined rotational symmetry $J M_J$. In general, the coefficients $c_{kl}$ in Eq.~\eqref{eq:divalent_wf} are determined from a CI (configuration-interaction) procedure involving diagonalization of the entire atomic Hamiltonian. For two-particle wavefunctions  constructed in this way (Eq.~\eqref{eq:divalent_wf_exp}), the reduced matrix elements of an ITO of rank $J$ can be written in terms of single-particle orbitals as~\cite{JohPlaSap95}
\begin{align} \label{eq:div_rmatel}
\bra rs &(J_F) ||T^{(J)}||mn(J_I)\ket 
  = \sqrt{(2J_I+1)(2J_F+1)} (-1)^{J} \sum_{\substack{{m\leq n} \\{r\leq s}}} 
  \eta_{rs} \eta_{mn} c_{rs} c_{mn} \\ \nonumber
  &\times \Bigg[ 
   (-1)^{j_r+j_s+J_I}\sixj{J}{J_I}{J_F}{j_s}{j_r}{j_m} 
      \bra r||t^{(J)}||m\ket \delta_{ns} 
    +(-1)^{j_r+j_n}\sixj{J}{J_I}{J_F}{j_s}{j_r}{j_n} 
      \bra r||t^{(J)}||n\ket \delta_{ms} \\ \nonumber
    &+(-1)^{J_F+J_I+1}\sixj{J}{J_I}{J_F}{j_r}{j_s}{j_m} 
      \bra s||t^{(J)}||m\ket \delta_{nr}  
    +(-1)^{j_r+j_n+J_F}\sixj{J}{J_I}{J_F}{j_r}{j_s}{j_n} 
      \bra s||t^{(J)}||n\ket \delta_{mr} 
  \Bigg] \;. 
\end{align}
Here $J_I$ and $J_F$ are the total angular momenta of the two-particle states $|j_r j_s\ket$ and $|j_m j_n\ket$ and $t^{(J)}(r)$ are the single-particle operators related to $T^{(J)}$ by $T^{(J)}=\sum_j t^{(J)}(r_j)$. For Rydberg excitations, we will employ a simplified ``independent particle approximation", where only one of the CI coefficients remains non-zero. 

\subsubsection{Polarizability}\label{subsubsec:magic_polariz}
Full interaction potential for the electrons in the electromagnetic field in the velocity gauge (also transverse or Coulomb gauge) is 
\begin{equation}\label{eq:hamiltonian}
V = - \sum_{j} \frac{\mathbf{A}_{\velg}(\mathbf{r}_j,t)\cdot\mathbf{p}_j}{c} 
	+ \sum_{j} \frac{A^2_{\velg}(\mathbf{r}_j,t)}{2c^2} \;. 
\end{equation}
Here $\mathbf{p}_j$ and $\mathbf{r}_j$ are the linear momentum operator and the coordinate of atomic electron $j$, and we used the Gaussian system of units. The vector potential of an optical lattice in the velocity gauge is 
\begin{equation}\label{eq:vec_pot}
\mathbf{A}_{\velg}(\mathbf{r}_j,t) = -\frac{2cF_0}{\omega} \hat{\epsilon} 
	\sin(k(Z+z_j)) \sin(\omega t) \;, 
\end{equation}
where we separated out the nuclear coordinate $Z$. In divalent atoms, there are two optically-active  electrons. Let us assume that one of the electrons is in the Rydberg state $|r\ket=|{n_{r}l_{r}j_{r}m_{r}}\ket$ and the other is in the ground state $|g\ket=|{n_{g}l_{g}j_{g}m_{g}}\ket$ of the remaining singly-charged core (by $m_g$ and $m_r$ we refer to the $z$-component of $j$). In the equations that follow, we will denote the quantum numbers $\{n_{j}l_{j}\}$ by $\gamma_{j}$ to simplify the notation. Then the energy shift due to the optical lattice  (second order in the field strength) for the two electron state $|{\gamma_{g}j_{g}}; {\gamma_{r}j_{r}}(JM)\ket$ becomes 
\begin{align}\label{eq:pert}
&\begin{aligned}
	\delta E_{gr}(\omega) = \frac{1}{4c^2} 
		&\sum_{g'r'} \frac{2\Delta E_{g'r'}}{\Delta E_{g'r'}^2-\omega^2} \\ 
			&\times \left \bra {\gamma_{g}j_{g}}; {\gamma_{r}j_{r}}(JM) \left |
				\sum_j \mathbf{A}_{\velg}(\mathbf{r}_j)\cdot\mathbf{p}_j \right | 
					{\gamma_{g'}j_{g'}}; {\gamma_{r'}j_{r'}}(J'M') \right \ket \\ 
			&\times \left \bra {\gamma_{g'}j_{g'}}; {\gamma_{r'}j_{r'}}(J'M') \left |
				\sum_{j'}  \mathbf{A}_{\velg}(\mathbf{r}_{j'})\cdot\mathbf{p}_{j'} \right | 
					{\gamma_{g}j_{g}}; {\gamma_{r}j_{r}}(JM) \right \ket \\ 
			&+ \frac{1}{2c^2} \sum_j \left\bra {\gamma_{g}j_{g}}; {\gamma_{r}j_{r}}(JM) \left|
				({A}_{\velg}(\mathbf{r}_j))^2 \right| {\gamma_{g}j_{g}}; {\gamma_{r}j_{r}}(JM) \right\ket \;, 
	\end{aligned}
\end{align}
where the summation is over the intermediate states $|g'r'\ket = |{\gamma_{g'}j_{g'}}; {\gamma_{r'}j_{r'}}(J'M')\ket$ and $\Delta E_{g'r'}$ are their energies. We will express $\delta E_{gr}(\omega)$ in terms of the conventional AC polarizability for a standing wave described by~\eqref{eq:vec_pot}, $\delta E_{gr}(\omega)=-\alpha_{gr}(\omega)F_0^2/4$. Eq.~\eqref{eq:pert} omits dynamic polarizability of the closed-shell core ({\it e.g.} Sr$^{++}$ for Sr). This contribution is negligibly small in the differential Stark shifts as its contribution is nearly identical for optically excited levels~\cite{DerPor11}. 

We will proceed as follows: First we will expand the two-particle operators in Eq.~\eqref{eq:pert} in ITOs. We will then assume the independent particle approximation and break up the two-particle matrix elements into linear combinations of single-particle matrix elements, while retaining the original overall rotational symmetry $J$. This results in an expression for $\alpha_{gr}(\omega)$ which looks like the sum of the polarizabilities for the individual one-electron systems: the ground and the Rydberg state electrons. The second order term that results from Eq.~\eqref{eq:pert} for the Rydberg electron is small compared to the $({A}_{\velg})^2$ term~\cite{TopDer13}, therefore we will ignore it. This is not the case for the ground state electron, and we will keep both the second order term and the $({A}_{\velg})^2$ term to evaluate its polarizability. Retaining $J$ of the original two-electron state will determine the extent to which the Rydberg electron polarizability contributes to $\alpha_{gr}(\omega)$. 

Now we focus on the first term in Eq.~\eqref{eq:pert} (which we also refer to as the second-order term). In order to cast this term into a tractable form, we expand the operators $\mathbf{A}_{\velg}(\mathbf{r}_j)\cdot\mathbf{p}_j$, Eq.~\eqref{eq:vec_pot}, in terms of ITOs. To this end, we first express $\mathbf{A}_{\velg}(\mathbf{r}_j)$ in ITOs: $\mathbf{A}_{\velg}(\mathbf{r}_j) = \sum_{K} \hat{\epsilon} \;S^{(K)}_{M_K=0}$, where $\hat{\epsilon}$ is the polarization vector and $S^{(K)}$ is an ITO of rank $K$. An explicit expression for $S^{(K)}$ is derived in the Appendix. $S^{(K)}$ are proportional to the spherical Bessel functions $j_K(2kr)$, and the $M_K=0$ limitation comes from the axial symmetry of Eq.~\eqref{eq:vec_pot}. 

Realizing that the momentum vector $\mathbf{p}_j$ is a tensor of rank 1, we can express $\mathbf{A}_{\velg}(\mathbf{r}_j)\cdot\mathbf{p}_j$ as an expansion in terms of composite tensor operator $B^{(L)}$
\begin{eqnarray}\label{eq:ap_ito}
\mathbf{A}_{\velg}(\mathbf{r}_j)\cdot\mathbf{p}_j  &=& \frac{F_0}{2} \sum_{\mu} (-1)^{\mu} \epsilon_{-\mu} 
			\sum_{L,M_L} {C}^{L\;M_L}_{1\;\mu\;K\;0} \; \{ S^{(K)} \otimes p^{(1)} \}^{(L)}_{M_L} \\
		&\equiv& \frac{F_0}{2} \sum_{L,\mu} (-1)^{\mu} \epsilon_{-\mu} B^{(L)}_{\mu}(K) \;,
\end{eqnarray}
where we defined $B^{(L)}_{\mu}(K)\equiv {C}^{L\;\mu}_{1\;\mu\;K\;0} \; \{ S^{(K)} \otimes p^{(1)} \}^{(L)}_{\mu}$ by realizing that the Clebsch-Gordan coefficient in~\eqref{eq:ap_ito} forces $M_L=\mu$. Notation $\{A^{(K_1)} \otimes B^{(K_2)}\}^{(L)}$ stands for a tensor of rank $L$ obtained by coupling tensors $A^{(K_1)}$ of rank $K_1$ and $B^{(K_2)}$ of rank $K_2$. 

Due to the small spatial extent of the ground state wavefunction of the singly charged-ion, we take only the $K=0$ term in the multipolar expansion of $\mathbf{A}_{\velg}(\mathbf{r}_j)$; this corresponds to the leading E1 multipole.  Furthermore, we assume the long wavelength approximation $kr\ll 1$ for the compact ground state, which yields the usual dipole approximation: $S^{(K)}\propto \delta_{K,0}$. For the Rydberg state, however, we will go beyond the long wavelength dipole approximation (see below). Keeping only the $K=0$ term for the ground state collapses $B^{(L)}_{\mu}(K)$ to the $L=1$ term alone, and the operator $\mathbf{A}_{\velg}(\mathbf{r}_j)\cdot\mathbf{p}_j$ becomes 
\begin{eqnarray}\label{eq:ap_ito_k0} 
\mathbf{A}_{\velg}(\mathbf{r}_j)\cdot\mathbf{p}_j 
	&=& \frac{F_0}{2} \sum_{\mu} (-1)^{\mu} \epsilon_{-\mu} B^{(1)}_{\mu}(0) \\ 
	&=& \frac{2c}{\omega} F_0 (\hat{\epsilon}\cdot\hat{p}) \;. 
\end{eqnarray}
Inserting Eq.~\eqref{eq:ap_ito_k0} into Eq.~\eqref{eq:pert} yields the product $B^{(1)}(0)\hat{R} B^{(1)}(0)$, where $\hat{R}$ denotes the resolvent operator (it is a scalar). Recoupling this product, we obtain ITOs $\{ B^{(1)}(0) \otimes \hat{R} B^{(1)}(0) \}^{(L')}$ of ranks $L'=0$, 1, and 2, that can be recognized as the conventional scalar, vector, and tensor polarizabilities. Since we focus on the $J=0$ states, only the scalar contribution to the polarizability remains, which is identified by $L'=0$ and its only component $\mu'=0$. We will evaluate this contribution later. 

We now turn to the second term in Eq.~\eqref{eq:pert} involving the expectation value of $(A_{\velg})^2$. In alkali-metal Rydberg atoms, this is the dominant term and it gives rise to the landscaping polarizability $\alpha^{\rm lsc}_{nlm_z}(\omega)$~\cite{TopDer13} reviewed in Sec.~\ref{subsec:magic_alkali}. Similar to the one-electron case in Eq.~\eqref{eq:alkali_tk}, we expand $(A_{\velg}(\mathbf{r}_j))^2$ in the single-particle operators $t^{(K)}(r_j)$ 
\begin{eqnarray}\label{eq:a2_ito}
({A}_{\velg}(\mathbf{r}_j))^2 = \frac{F_0^2}{4} \sum_K t^{(K)}_{M_K=0}(r_j) \;, 
\end{eqnarray}
which are related to the many-body operator $T^{(K)}$ through $T^{(K)}= \sum_j t^{(K)}(r_j)$. We use the Wigner-Eckart theorem to integrate over the magnetic quantum numbers so that the matrix element in the second term in Eq.~\eqref{eq:pert} is expressed in terms of the reduced matrix element 
\begin{eqnarray}\label{eq:a2_wig_eck}
\bra \gamma_{g}\gamma_{r} JM|T^{(K)}_{0} |\gamma_{g}\gamma_{r} JM\ket 
	&=& (-1)^{J-M} \threej{J}{K}{J}{-M}{0}{M} 
		\bra\gamma_{g}\gamma_{r} J ||T^{(K)}||\gamma_{g}\gamma_{r} J\ket \;.
\end{eqnarray}
Because the $(A_{\velg})^2$ term is the expectation value in the state $|\gamma_g \gamma_r JM\ket$, the Wigner-Eckart theorem limits $K$ to $2J$. Moreover, since we are only interested in the $J=0$ states, $K=0$. This restricts the representation of $(A_{\velg})^2$ in term of ITO in Eq.~\eqref{eq:a2_ito} to the $t^{(0)}_{0}(r_j)$ term alone. 

In evaluating the many-body matrix elements, we will assume the independent electron approximation. In this approximation, the reduced matrix element in Eq.~\eqref{eq:a2_wig_eck} can be broken up into reduced matrix elements involving only the one-electron orbitals using Eq.~\eqref{eq:div_rmatel} 
\begin{align} 
\bra j_g j_r(J) ||T^{(K)} &||j_g j_r(J)\ket 
  = (2J+1) (-1)^{K+j_g+j_r+J}  \nonumber \\
  &\times 
  \Bigg[ 
   \sixj{K}{J}{J}{j_r}{j_g}{j_g} 
      \bra g||t^{(K)}||g\ket 
  	+\sixj{K}{J}{J}{j_r}{j_g}{j_r}  
      \bra r||t^{(K)}||r\ket 
  \Bigg] \;. 
\end{align}
Two of the terms in Eq.~\eqref{eq:div_rmatel} dropped out because we assume that $|g\ket$ and $|r\ket$ are two distinct one-electron states, hence $\delta_{gr}=0$. 

We now put together the second-order term and the $(A_{\velg})^2$ term expressed in terms of one-particle reduced matrix elements to obtain an expression for $\alpha_{gr}(\omega)$ following from Eq.~\eqref{eq:pert}. We are particularly interested in the states $|n_{g}s_{1/2}n_{r}s_{1/2}(^1S_0)\ket$ and $|n_{g}s_{1/2}n_{r}p_{1/2}(^3P_0)\ket$. In the independent electron approximation, the energy of the two-electron state $|g'r'\ket$ in can be separated into individual contributions from the one-electron states: $E_{g'r'} \approx E_{g'} + E_{r'}$. These approximations simplify Eq.~\eqref{eq:pert} greatly, and the dynamic polarizability $\alpha(\omega)$ separates into individual contributions from the ground and the Rydberg states 
\begin{align}
&\begin{aligned}\label{eq:pert2}
	\alpha_{gr}(\omega) = 
		-\frac{1}{\omega^2}
		&\sum_{\substack{{g'\neq g}}} 
			\frac{2\Delta E_{g'}}{\Delta E_{g'}^2-\omega^2} 
			\sum_{\substack{{L,J',\mu}}} (-1)^{\mu} \epsilon_{-\mu} 
			\threej{0}{L}{J'}{0}{\mu}{M'}\sixj{L}{J'}{0}{\sfrac{1}{2}}{\sfrac{1}{2}}{\sfrac{1}{2}} 
				\sixj{L}{0}{J'}{\sfrac{1}{2}}{\sfrac{1}{2}}{\sfrac{1}{2}} \\ 
			&\times
			\bra {n_{g}l_{g} \sfrac{1}{2}} ||
				b^{(L)}_{\mu}(r) || {n_{g'}l_{g'} \sfrac{1}{2}}\ket 
			\bra {n_{g'}l_{g'} \sfrac{1}{2}} ||
				b^{(L)}_{\mu}(r) || {n_{g}l_{g} \sfrac{1}{2}}\ket \\ 
		& 
			-\frac{1}{\omega^2}
			\sum_{r'\neq r} 
			\frac{2\Delta E_{r'}}{\Delta E_{r'}^2-\omega^2} 
			\sum_{\substack{{L,J',\mu}}} (-1)^{\mu} \epsilon_{-\mu} 
			\threej{0}{L}{J'}{0}{\mu}{M'} \sixj{L}{J'}{0}{\sfrac{1}{2}}{\sfrac{1}{2}}{\sfrac{1}{2}} 
				\sixj{L}{0}{J'}{\sfrac{1}{2}}{\sfrac{1}{2}}{\sfrac{1}{2}}  \\ 
			&\times 
			\bra {n_{r}l_{r} \sfrac{1}{2}} ||
				b^{(L)}_{\mu}(r) || {n_{r'}l_{r'} \sfrac{1}{2}}\ket 
			\bra {n_{r'}l_{r'} \sfrac{1}{2}} ||
				b^{(L)}_{\mu}(r) || {n_{r}l_{r} \sfrac{1}{2}}\ket \\ 
			& 
					-\frac{1}{\omega^2}
					\sum_{K} \threej{0}{K}{0}{0}{0}{0} 
					\sixj{K}{0}{0}{\sfrac{1}{2}}{\sfrac{1}{2}}{\sfrac{1}{2}}
					\bra {n_{g}l_{g} \sfrac{1}{2}} ||t^{(K)}(r) ||{n_{g}l_{g} \sfrac{1}{2}} \ket \\
			&	
					-\frac{1}{\omega^2}
					\sum_{K} \threej{0}{K}{0}{0}{0}{0} 
					\sixj{K}{0}{0}{\sfrac{1}{2}}{\sfrac{1}{2}}{\sfrac{1}{2}}					
					\bra {n_{r}l_{r} \sfrac{1}{2}} ||t^{(K)}(r) ||{n_{r}l_{r} \sfrac{1}{2}} \ket \;. 
	\end{aligned}
\end{align}
Here we have expressed $B^{(L)}$ in terms of the one-electron operators: $B^{(L)}=\sum_j b^{(L)}(r_j)$. To split the second-order term in~\eqref{eq:pert} into individual contributions, we used the fact that the main contribution to $\alpha_{gr}(\omega)$ for a given state comes from states that are nearby in energy, and the overlaps between the ground and Rydberg states are small, {\it i.e.} $\bra \gamma_{g} j_{g} |b^{(L)}_{\mu}(r)| \gamma_{r'} j_{r'}\ket \ll \bra \gamma_{g} j_{g} |b^{(L)}_{\mu}(r)| \gamma_{g'} j_{g'}\ket$ and $\bra \gamma_{r} j_{r} |b^{(L)}_{\mu}(r)| \gamma_{g'} j_{g'}\ket \ll \bra \gamma_{r} j_{r} |b^{(L)}_{\mu}(r)| \gamma_{r'} j_{r'}\ket$. Therefore we neglected terms involving $\bra \gamma_{g} j_{g} |t^{(K)}| \gamma_{r'} j_{r'}\ket$ and $\bra \gamma_{r} j_{r}|t^{(K)}| \gamma_{g'} j_{g'}\ket$ and arrived at the first two terms in Eq.~\eqref{eq:pert2}.  

The first and the thirds terms can be consolidated together to give the ground state polarizability for the singly-charged ionic core in the velocity gauge. For the Sr atom, this would be the polarizability of the $5s$ state of the Sr$^+$ ion, $\alpha_{5s}^{{\rm Sr}^+}(\omega)$. Because of the gauge invariance, $\alpha_{5s}^{{\rm Sr}^+}(\omega)$ can be calculated either in the velocity or the length gauges, and we will evaluate $\alpha_{5s}^{{\rm Sr}^+}(\omega)$ using the length gauge below. The second and the fourth terms in~\eqref{eq:pert2} result from the Rydberg electron. The 3-$j$ symbol in the last term, which came from integrating over the magnetic quantum numbers using the Wigner-Eckart theorem in~\eqref{eq:a2_wig_eck}, collapses the sum $\sum_K t^{(K)}$ to the $K=0$ term alone. Furthermore, $L=1$ and $M_L=\mu$ forces $M'=-\mu$ in the second term. With these simplifications and using the closed form expressions for the 3-$j$ and 6-$j$ symbols which appear in~\eqref{eq:pert2}, the dynamic polarizability can be rewritten as 
\begin{align}
&\begin{aligned}\label{eq:pert_alp}
	\alpha_{gr}(\omega) = \alpha_{5s}^{{\rm Sr}^+}(\omega) 
		-\frac{1}{\omega^2} \sum_{\substack{{L,\mu} \\{r'\neq r}}} &(-1)^{\mu} \epsilon_{-\mu}
			\frac{2\Delta E_{r'}}{\Delta E_{r'}^2-\omega^2} 
			\frac{(-1)^{L-\mu} \delta_{L,J'} \delta_{\mu,-M'}}{2(2L+1)^{3/2}} \\
			&\times |\bra {n_{r}l_{r} \sfrac{1}{2}} ||
				b^{(L)}_{\mu}(r) || {n_{r'}l_{r'} \sfrac{1}{2}}\ket |^2 
			-	\frac{1}{\omega^2\sqrt{2}}			
					\bra {n_{r}l_{r} \sfrac{1}{2}} ||t^{(0)}(r) ||{n_{r}l_{r} \sfrac{1}{2}} \ket \;. 
	\end{aligned}
\end{align}

The dominant contribution for the Rydberg electron comes from the last term as discussed in~\cite{TopDerGauge13}. The correction to the $A_{\velg}^2$ term is given by the second term in~\eqref{eq:pert3} and is negligibly small for Rydberg states as demonstrated in~\cite{TopDerGauge13}. The $A_{\velg}^2$ term is what gives rise to the landscaping polarizability in alkali metal atoms, and is proportional to $\cos(2kz)$. Therefore the total polarizability of the $J=0$ divalent Rydberg atom can be expressed as 
\begin{align}
&\begin{aligned}\label{eq:pert3}
	\alpha^{J=0}_{n_{g}l_{g};n_{r}l_{r}}(\omega) = \alpha_{\rm ion}(\omega) 
		+ \alpha^{{\rm lsc},J=0}_{n_{r}l_{r}}(\omega) 
		+ \alpha_{\rm core}(\omega) 
		+ \alpha_{\rm cv}(\omega) 
		\;, 
	\end{aligned}
\end{align}
where $\alpha_{\rm ion}(\omega)$ is the polarizability of the residual ion ({\it e.g.} Sr$^+$) and $\alpha^{{\rm lsc},J=0}_{n_{r}l_{r}}(\omega)$ is the contribution from the Rydberg landscaping polarizability to the total polarizability of the $J=0$ two-electron state. The notation $\alpha^{J=0}_{n_{g}l_{g};n_{r}l_{r}}(\omega)$ refers to the total polarizability defined in~\eqref{eq:pert2} and~\eqref{eq:pert_alp} ($\alpha_{gr}(\omega)$). The polarizability $\alpha_{\rm core}(\omega)$ comes from the contributions from core-excited states of doubly ionized atom ({\it e.g.} Sr$^{++}$) to the total polarizability~\cite{DerPor11}. We will neglect this term because it is almost identical for both valence levels and it vanishes when only the differential contribution is considered. We also neglect $\alpha_{\rm cv}(\omega)$, which is a small term counteracting $\alpha_{\rm core}(\omega)$. It arises from excitations to occupied valence orbitals and is much smaller than $\alpha_{\rm core}(\omega)$~\cite{DerPor11}. 

The $J=0$ contribution $\alpha^{{\rm lsc},J=0}_{n_{r}l_{r}}(\omega)$ can be explicitly written as 
\begin{eqnarray}\label{eq:alp_rlsc}
\alpha^{{\rm lsc},J=0}_{n_{r}l_{r}}(\omega) = -	\frac{1}{\omega^2\sqrt{2}}			
		\bra {n_{r}l_{r} \sfrac{1}{2}} ||t^{(0)}(r) ||{n_{r}l_{r} \sfrac{1}{2}} \ket \;. 
\end{eqnarray}
Therefore the consequence of the overall rotational symmetry of the many-body state is that only the $K=0$ term in $(A_{\velg})^2$ (Eq.~\eqref{eq:a2_ito}) contributes to the total polarizability. To calculate the reduced matrix element in $\alpha^{{\rm lsc},J=0}_{n_{r}l_{r}}(\omega)$, we use the same expansion used for the alkali-metal atoms in~\eqref{eq:alkali_tk} 
\begin{eqnarray}
\bra {n_{r}l_{r} \sfrac{1}{2}} ||t^{(0)}(r) ||{n_{r}l_{r} \sfrac{1}{2}} \ket 
	&=& \bra {n_{r}l_{r} \sfrac{1}{2}}||C^{(0)}(\hat{r})|| {n_{r}l_{r} \sfrac{1}{2}}\ket 
			\int_{0}^{\infty} P_{n_r l_r}^2(r) j_0(2kr) \;dr \\
	&=& \sqrt{2} \int_{0}^{\infty} P_{n_r l_r}^2(r) j_0(2kr) \;dr \;. 
\end{eqnarray}
Thus, 
\begin{eqnarray}\label{eq:alp_rlsc_integral}
\alpha^{{\rm lsc},J=0}_{n_{r}l_{r}}(\omega) = -\alpha_{\rm e}(\omega) 
		\int_{0}^{\infty} P_{n_r l_r}^2(r) j_0(2kr) \;dr \;. 
\end{eqnarray}

For the 5s ground state of the Sr$^+$ ion, the trapping potential reads $U_{\rm Sr^+} = -(F^2_0 /4)\alpha_{\rm ion}(\omega) \sin^2(kZ)$ where the dynamic polarizability fo the Sr$^+$ ion is given by 
\begin{equation}
\alpha_{\rm ion}(\omega) = \sum_{j} \frac{E_{5s}-E_{j}}{(E_{5s}-E_{j})^2 -\omega^2} 
	|\bra \psi_{\rm 5s} | \mathbf{D}|\psi_{j} \ket|^2 \;,
\end{equation}
where $\mathbf{D}$ is the electric dipole operator and $E_j$ are the ionic energy levels. We evaluate $\alpha_{\rm ion}(\omega)$ using a high-accuracy method detailed in Ref.~\cite{DerJohSaf99}.

\section{Numerical results for Sr} \label{sec:results}
This section is organized as follows: in Sec.~\ref{subsec:ion_rydberg}, we evaluate the $5s$ ground state polarizability for the Sr$^+$ ion and the $J=0$ landscaping polarizabilities for a few Rydberg states and discuss their general features. To find magic wavelengths at which clock-to-Rydberg state transition frequencies do not change, we then calculate the dynamic polarizabilities for the $5s^2(^1S_0)$ and $5s5p(^3P_0)$ clock states of Sr in Sec.~\ref{subsec:clock}. Finally in Sec.~\ref{subsec:divalent_states}, we evaluate the total divalent Rydberg state polarizabilities by combining these as described in Sec.~\ref{subsec:magic_divalent} to find the magic wavelengths. 

\subsection{Residual ion and the Rydberg electron}\label{subsec:ion_rydberg}
\begin{figure}[h!tb]
	\begin{center}
     \resizebox{85mm}{!}{\includegraphics{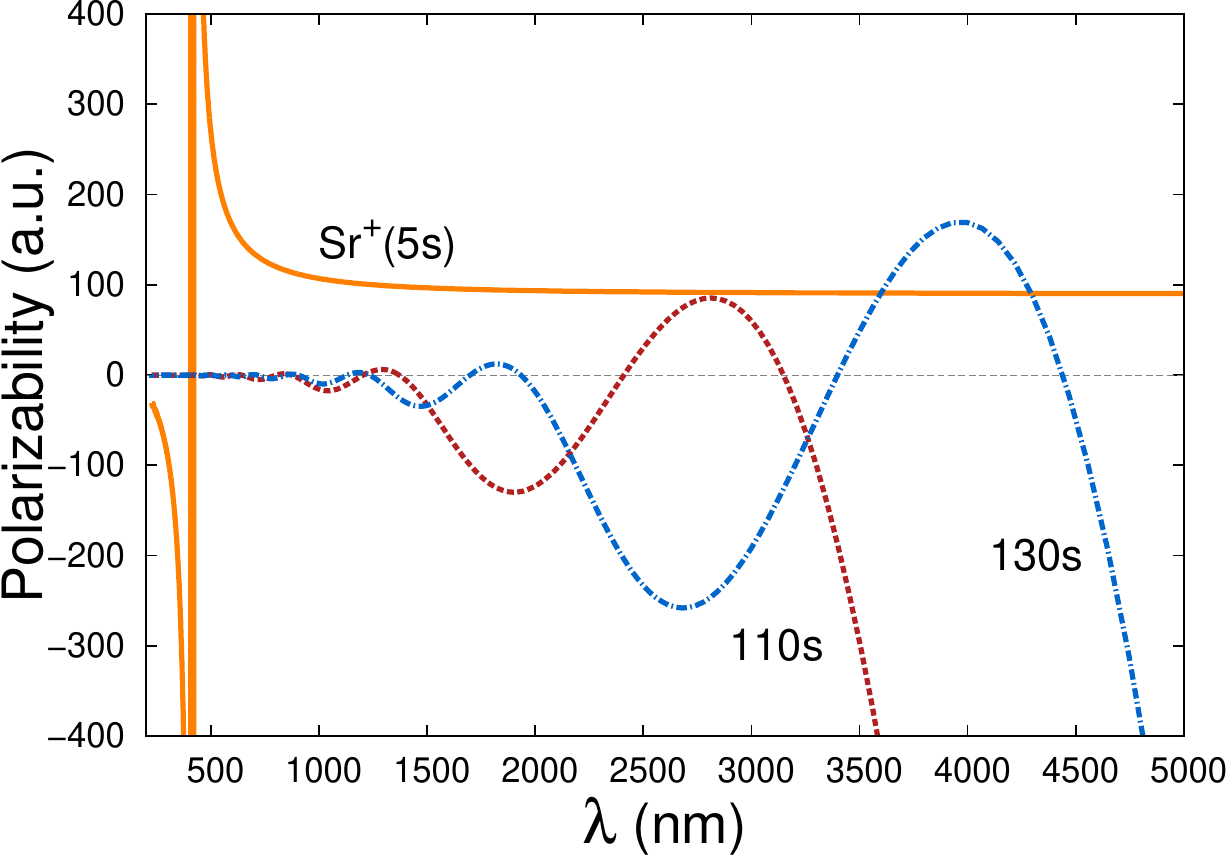}} 
  \end{center}
  \caption{(Color online) Polarizability of the Sr$^+$ ion in the 5s ground state 
  (solid orange), and the landscaping polarizabilities $\alpha^{{\rm lsc},J=0}_{ns}(\lambda)$ 
  for the Rydberg electron in the $110s$ (dashed maroon) and $130s$ (dot-dashed blue) states of 
  the Sr atom. 
  }
  \label{fig:srplus_ns}
\end{figure}

We start by calculating the $5s$ ground state polarizability for the Sr$^+$ ion and the landscaping polarizabilities for a few $n$s Rydberg states. The main features of $n$p Rydberg states are essentially the same as discussed in Sec.~\ref{subsubsec:magic_polariz}. These individual  polarizabilities can be added to obtain the total polarizability of the Sr atom in the Rydberg state 5s$n$s($^1S_0$) as we demonstrated in the previous section. Fig.~\ref{fig:srplus_ns} shows the polarizability for the Sr$^+$ ion in the 5s ground state and the landscaping polarizabilities $\alpha^{{\rm lsc},J=0}_{n_{r}l_{r}}(\lambda)$ for the Rydberg electron in the 5s$n$s($^1S_0$) state of Sr for $n=110$ and 130. The main feature of $\alpha_{\rm ion}$ is that it converges to its static value after $\lambda\sim$1000 nm at $\sim 93$ a.u.. On the other hand, the Rydberg landscaping polarizabilities start out essentially at zero at small $\lambda$, and oscillate with increasing amplitude towards larger $\lambda$, before the free electron character of the polarizability takes over and $\alpha^{{\rm lsc},J=0}_{ns}(\omega)$ drops off as $\alpha_{\rm e}(\omega)$. 

The radial wave functions $P_{nl}(r)$ needed to evaluate $\alpha^{{\rm lsc},J=0}_{n_{r}l_{r}}(\omega)$ are computed by numerical integration of the time-independent  Schr\"odinger equation using a model for the ionic core seen by the Rydberg electron in state $|nljm_j\ket$. Both in $5sns$($^1S_0$) and $5snp$($^3P_0$) singly excited Rydberg states of Sr, the Rydberg electron moves under the influence of the Sr$^+$ potential. We model this potential as 
\begin{equation}
V(r) = -\frac{1}{r}-\frac{(Z_a-1){\rm e}^{-ar}}{r} +b{\rm e}^{-cr} \;,
\end{equation}
where $Z_a$ is the atomic number (38 for Sr) and $a$, $b$ and $c$ are fitting parameters, which depend on the particular symmetry of the many-body Rydberg state. The coefficients $a$, $b$ and $c$ for Sr are listed in Ref.~\cite{Millen11} for a variety of $LSJ$ symmetries. For the $5sns$($^1S_0$) states of Sr, we use $a=3.762$, $b=-6.33$ and $c=1.07$, and for the $5snp$($^3P_0$) states we use $a=3.45$, $b=-6.02$ and $c=1.07$. 

\subsection{The clock states}\label{subsec:clock}
In order to determine magic wavelengths, we need to calculate the dynamic polarizabilities for the 5s$^2$($^1S_0$) and 5s5p($^3P_0$) states of Sr. We use the relativistic formulation of the second order perturbation theory~\cite{JohPlaSap95}, 
\begin{align}
\begin{aligned} 
\alpha^{J=0}_{\gamma}(\omega) = 
	-\frac{1}{\sqrt{3}} \sum_{\gamma'} &\sixj{1}{1}{0}{0}{0}{1}
		\frac{2(E_{\gamma J=0}-E_{\gamma'J'=1})}{(E_{nJ=0}-E_{\gamma'J'=1})^2-\omega^2} \\ 
			&\times \bra \gamma (J=0)||D||\gamma'(J'=1)\ket \bra \gamma'(J'=1)||D||\gamma (J=0)\ket \;. 
\end{aligned}	
\end{align}
In the above expression, we have specifically chosen $J=0$ and $|\gamma\ket$ refers to the states $|5s^2(^1S_0)\ket$ and $|5s5p(^3P_0)\ket$. We use high-accuracy values for the reduced matrix elements $\bra \gamma (J=0)||D||\gamma'J'\ket$ and experimental energies $E_{\gamma J}$ tabulated in Ref.~\cite{SafPorSaf13}. 

\begin{figure}[h!tb]
	\begin{center}
     \resizebox{85mm}{!}{\includegraphics{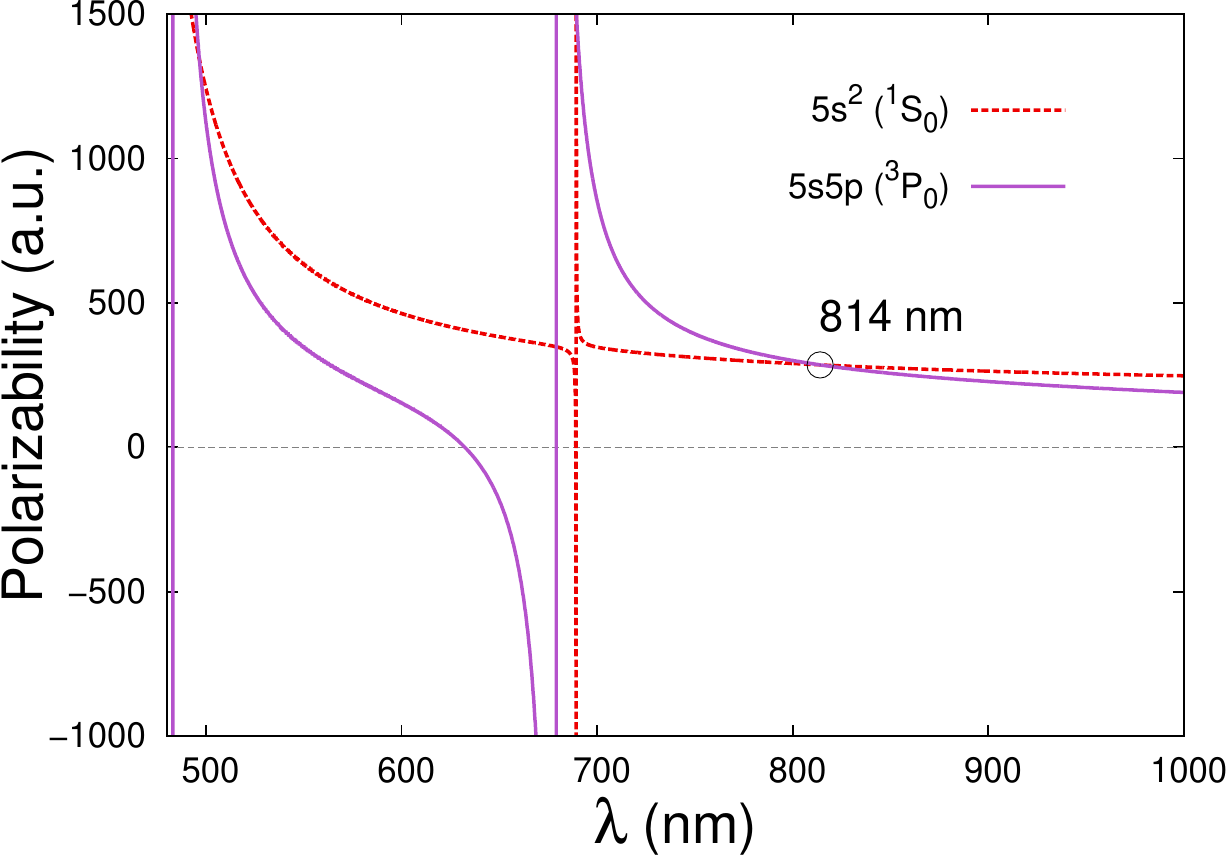}} 
  \end{center}
  \caption{(Color online) Dynamic polarizabilities for the 5s$^2$($^1S_0$) (dashed red) 
  ground and the 5s5p($^3P_0$) excited clock states (solid purple) of Sr. The 
  magic wavelength where the two polarizabilities are equal is marked by an open circle at 814 nm. 
  }
  \label{fig:ground_1S_3P}
\end{figure}

We can reproduce the magic wavelength commonly used for state-insensitive trapping Sr in optical lattice clocks by plotting the polarizabilities for the ground and the lowest lying clock states. This serves as a check of our calculations. In Fig.~\ref{fig:ground_1S_3P}, we plot the polarizabilities for the 5s$^2$($^1S_0$) ground state and the 5s5p($^3P_0$) clock state of Sr. We find that the magic wavelength for the 5s$^2$($^1S_0$) ground and the 5s5p($^3P_0$) clock states is at 814.0 nm, which perfectly matches the spectroscopically measured experimental value of 813.4 nm~\cite{TakKat03,TakHonHig05}. Therefore, we reproduce the ac polarizabilities of the 5s$^2$($^1S_0$) and 5s5p($^3P_0$) states. 

\subsection{Divalent Rydberg states}\label{subsec:divalent_states}
Now we construct the total divalent Rydberg state polarizabilities out of the ionic polarizability $\alpha_{\rm ion}(\omega)$ and the Rydberg landscaping polarizabilities $\alpha^{{\rm lsc},J=0}_{n_{r}l_{r}}(\omega)$. In this paper, we are interested in the $J=0$ Rydberg state of Sr in which one of the valence electrons is a spectator whereas the other is in a high $n$ level. Particularly, we focus on states such as 5s$n$s($^1S_0$) and 5s$n$p($^3P_0$). We have already demonstrated that the scalar polarizability for the $^1S_0$ Rydberg states can be evaluated by adding the polarizability of the Sr$^+$ ion $\alpha_{\rm ion}(\omega)$ and the landscaping polarizability $\alpha^{{\rm lsc},J=0}_{ns}(\omega)$ of the Rydberg electron (Eq.~\eqref{eq:pert3}). For the $^3P_0$ states, however, only part of the Rydberg landscaping polarizability contributes: 
\begin{eqnarray}
\alpha^{J=0}_{5sns}(\omega) &=& \alpha_{\rm ion}(\omega) 
	+ \alpha^{{\rm lsc},J=0}_{ns}(\omega) \label{eq:5sns} \\
\alpha^{J=0}_{5snp}(\omega) &=& \alpha_{\rm ion}(\omega) 
	+ \alpha^{{\rm lsc},J=0}_{np}(\omega) \;. \label{eq:5snp}
\end{eqnarray}

\begin{figure}[h!tb]
	\begin{center} 
		 \resizebox{85mm}{!}{\includegraphics{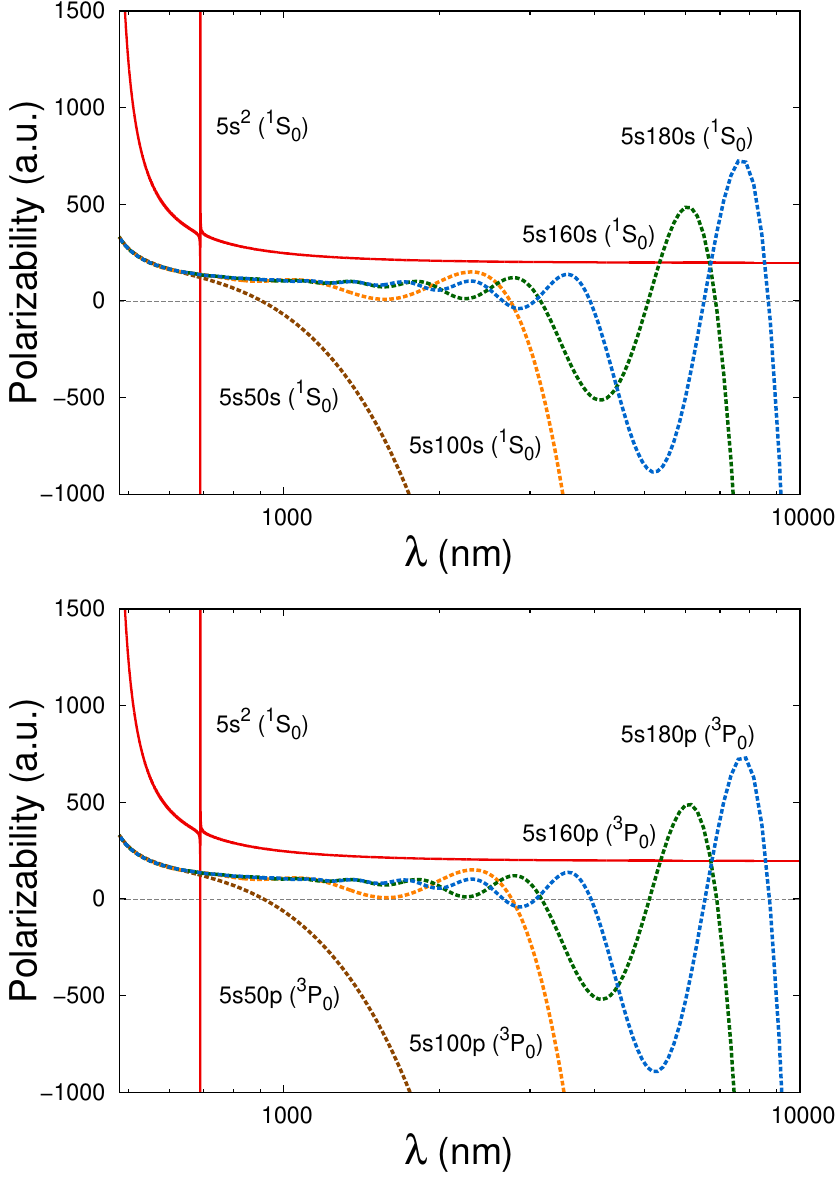}} 
     \end{center}
     \caption{(Color online) Total polarizabilities  
     for the 5s$n$s($^1S_0$) (upper panel) and 5s$n$p($^3P_0$) (lower panel) Rydberg states 
     of Sr for $n=50$ (dashed brown), 100 (orange), 160 (green) and 180 (blue) plotted with 
     the ground state 5s$^2$($^1S_0$) polarizability (solid red). 
     }
     \label{fig:ground_1S_ryd}
\end{figure}

In Fig.~\ref{fig:ground_1S_ryd}, we plot the dynamic polarizability of the 5s$^2$($^1S_0$) Sr ground state (solid purple) with the Rydberg state polarizabilities, $\alpha^{J=0}_{5sns}(\lambda)$ and $\alpha^{J=0}_{5snp}(\lambda)$. The upper panel of Fig.~\ref{fig:ground_1S_ryd} contains Rydberg states $|r\ket=$$|$5s$n$s($^1S_0$)$\ket$ and the lower panel contains $|r\ket=$$|$5s$n$p($^3P_0$)$\ket$ states for four principle quantum numbers: $n=50$, 100, 160 and 180. 

\begin{figure}[h!tb]
	\begin{center} 
		 \resizebox{85mm}{!}{\includegraphics{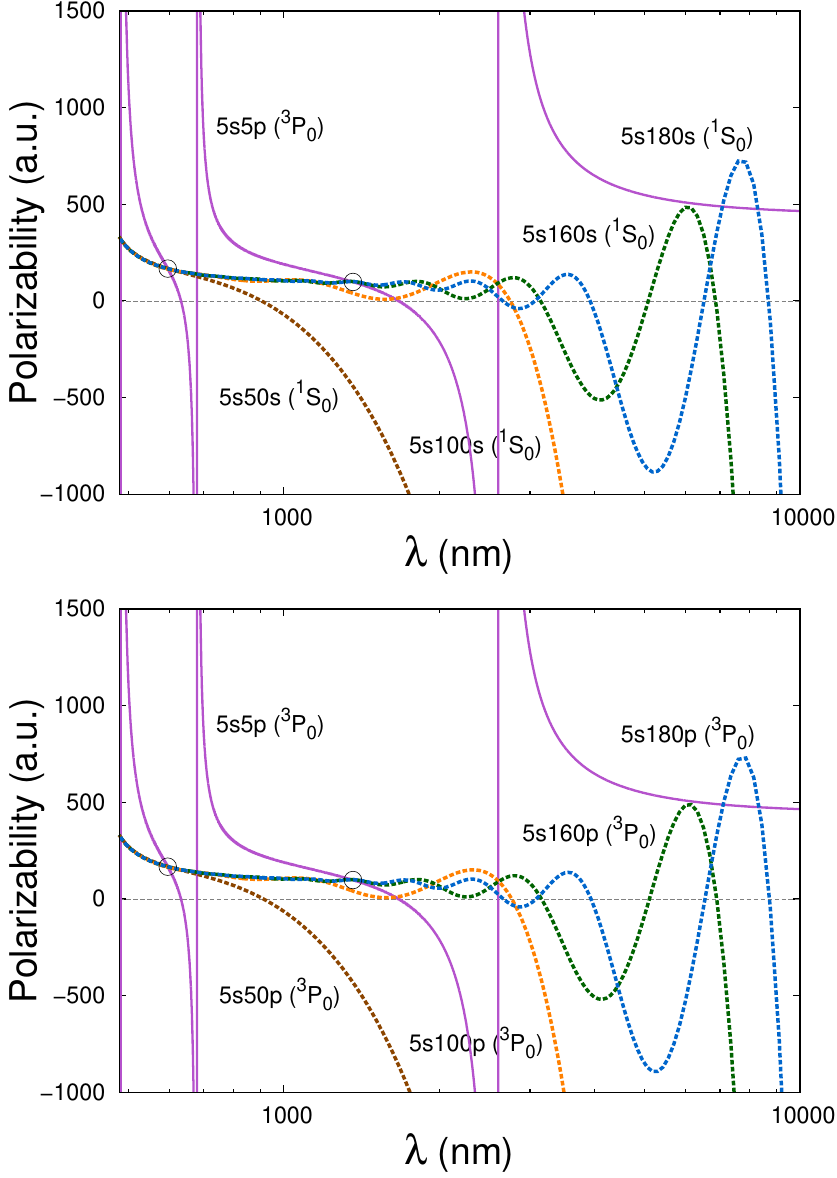}} 
     \end{center}
     \caption{(Color online) Same as Fig.~\ref{fig:ground_1S_ryd} except the ground state 
     polarizability is replaced by the polarizability of the upper clock state 5s5p($^3P_0$) 
     (solid purple). Two special points at which the 5s5p($^3P_0$) state polarizability matches 
     those of the Rydberg states in the high-$n$ limit are marked by open circles. The ``universal" 
     magic wavelengths corresponding to these points are at 596 nm and 1362 nm. 
     }
     \label{fig:ground_3P_ryd}
\end{figure}

For Rydberg states with $n=160$ and 180, the figure shows two wavelengths for simultaneous magic trapping of the ground state and the Rydberg states in lattices with wavelengths $\lambda >$ 5000 nm. It is worth emphasizing that the total polarizabilities for divalent Rydberg states are different for different $5sns$($^1S_0$) and $5snp$($^3P_0$) configurations (see Eq.~\eqref{eq:alp_rlsc}). However, both are expressed in terms of the $t^{(0)}\propto j_0(2kz)$ alone. Because of this, the Rydberg state polarizabilities in the upper and the lower panels in Fig.~\ref{fig:ground_1S_ryd} are almost exactly the same. This is in stark contrast with the case of $s$- and $p$-states in alkali-metal atoms. These magic wavelengths arise due to the intensity landscape modulation and are entirely new when compared with those reported in Ref.~\cite{MukMilNat11}. 

\begin{center}
\begin{figure}[h!tb]
\captionof{table}{Magic wavelengths (nm) for Rydberg states and the 5s$^2$($^1S_0$) and 5s5p($^3P0$) clock states of Sr seen in Figures~\ref{fig:ground_1S_ryd} and~\ref{fig:ground_3P_ryd}. Only wavelengths in the CO$_2$ laser band are tabulated. \\}
\begin{tabular}{ C{1.4cm} C{1.8cm} C{1.8cm} C{1.8cm} C{1.8cm}}
\hline\hline
 &\multicolumn{2}{c}{$^1S_0$}&\multicolumn{2}{c}{$^3P_0$}\\
\cline{2-5}
     & $5s160s$ & $5s180s$    & $5s160p$ & $5s180p$ \\
\hline
$5s^2(^1S_0)$   & 5347 \newline 6686 
	              & 6725 \newline 8542 
	              & 5372 \newline 6719 
	              & 6754 \newline 8578  \\ 
\hline
$5s5p(^3P_0)$   & 6076 
	              & 7078 \newline 8285 
	              & 6157 
	              & 7101 \newline 8326  \\ 
\hline
\end{tabular}
\label{table:magic_wlen}
\end{figure}
\end{center}

On the other hand, Fig.~\ref{fig:ground_3P_ryd} shows the same Rydberg state polarizabilities with the upper clock state 5s5p($^3P_0$) (solid purple). This time there are four magic wavelengths for $n=180$: two below 1000 nm (not counting the one right on the resonance) and two above 5000 nm. For the 5s160s($^1S_0$) and 5s160p($^3P_0$) there is only one wavelength allowing for the magic trapping condition. Because $\alpha^{{\rm lsc},J=0}_{ns}(\omega)$ is almost zero below 1000 nm, $\alpha^{J=0}_{5sns}(\omega)$ is essentially same as $\alpha_{\rm ion}(\omega)$, and in the long $\lambda$ region it is dominated by the Rydberg landscaping polarizability. Because $\bra\cos(2kz)\ket\rightarrow 0$ as $n\rightarrow \infty$ for a given value of $\lambda$, at the points marked by open circles in Fig.~\ref{fig:ground_1S_ryd}, the magic trapping condition is satisfied between the 5s160s($^1S_0$) and 5s160p($^3P_0$) states and the Rydberg states, regardless of the value of $n$ in the Rydberg states. In this sense, the magic wavelengths at 596 nm and 1362 nm are universal as they do not depend on $n$ of the Rydberg electron. Table~\ref{table:magic_wlen} lists the magic wavelengths seen in Figures~\ref{fig:ground_1S_ryd} and~\ref{fig:ground_3P_ryd} in the $\lambda >5000$ nm region. Whereas these wavelengths depend on the specific $n$ quantum number of the Rydberg states 5s$n$s($^1S_0$) and 5s$n$p($^3P_0$), the two universal wavelengths at 596 nm and 1362 nm are independent of $n$ and are essentially the same for all Rydberg states beyond $n\sim 160$. 

\section{Conclusion}\label{sec:conclusions}
The second (non-Rydberg) valence electron of divalent atoms sizably contributes to the  trapping potential and greatly simplifies the trapping of Rydberg states of divalent atoms. We have shown that although the valence and the Rydberg electron polarizabilities individually add to make up the polarizability in a $5sns$($^1S_0$) state, the situation is more complicated in the general case. For example, in a $5snp$($^3P_0$) state, only part of the Rydberg electron polarizability contributes to the overall polarizability of the divalent atom. The contribution from the Rydberg electron to the total polarizability of the divalent system is the landscaping polarizability $\alpha^{{\rm lsc},J=0}_{n_{r}l_{r}}(\omega)$, which is only one of the terms in the landscaping polarizability of the one-electron Rydberg state when expressed in terms of ITOs. Like the landscaping polarizability of alkali-metal atoms, $\alpha^{{\rm lsc},J=0}_{n_{r}l_{r}}(\omega)$ depends on the parameter $a_0n^2/\lambda$ and vanishes in the limit $\lambda\rightarrow 0$ (or $n\rightarrow \infty$). As a result, if it were not for the ac polarizability of the residual ion, the atom would become untrappable. In the opposite limit $a_0 n^2/\lambda \ll 1$, $\alpha^{{\rm lsc},J=0}_{n_{r}l_{r}}(\omega)$ approaches the free electron value. 

We explored the possibility of magic trapping of Rydberg states of divalent atoms (specifically Sr) with the 5s$^2$($^1S_0$) ground and 5s5p($^3P_0$) clock states. We find that these conditions can be satisfied at various lattice laser wavelengths in the IR region of the spectrum. The specific values of the magic wavelengths depend on the principal quantum number of the Rydberg state as the total dynamic polarizability in this region is dominated by the landscaping polarizability of the Rydberg electron. On the other hand, we also identified two wavelengths (596 nm and 1362 nm) at which the magic trapping can be attained for the upper clock state and the Rydberg states, whose values are independent of $n$ in the high-$n$ limit. This is due to the fact that the polarizability of the Sr Rydberg states are dominated by that of the  Sr$^+$ ion in the short $\lambda$ region, where these universal magic conditions occur. 

\section{Acknowledgements}
This work was supported by the National Science Foundation (NSF) Grant No. PHY-1212482. 

\section{Appendix} 
Here, we derive an explicit expression for the $S^{(K)}_{M_K}$ tensors, which were used to expand $\mathbf{A}_{\velg}(\mathbf{r}_j)$ in terms of ITOs. For a one-dimensional optical lattice formed by superposing two counter-propagating laser beams in the $z$-direction, the vector potential is given by~\eqref{eq:vec_pot}. It is proportional to $\sin[k(Z+z)]$. We begin by expressing this in complex exponentials: 
\begin{eqnarray}\label{eq:appdx_sine}
\sin[k(Z+z)] = \frac{1}{2i} (e^{ikZ} e^{ikz} - e^{-ikZ} e^{-ikz}) \;. 
\end{eqnarray}
The complex exponentials themselves can be expanded in terms of the spherical Bessel functions $j_K(kr)$: 
\begin{eqnarray}
e^{ikz} = \sum_{K=0}^{\infty} (2K+1) i^K j_K(kr) C^{(K)}_{M_K=0}(\hat{r}) \;. 
\end{eqnarray}
Substituting this expansion in Eq.~\eqref{eq:appdx_sine} and combining the right hand side term by term, we obtain 
\begin{eqnarray}\label{eq:appdx_dummy}
\sin[k(Z+z)] = \sum_{K=0}^{\infty} (2K+1) j_K(kr) C^{(K)}_{M_K=0}(\hat{r}) 
	\left[ \frac{i^K e^{ikZ}}{2i} +{\rm c.c.}  \right ] \;. 
\end{eqnarray}
Realizing that $i^K=\exp(-i\pi K/2)$, we finally obtain an expression for $S^{(K)}_{M_K=0}$: 
\begin{eqnarray}
S^{(K)}_{M_K=0} = -\frac{2cF_0}{\omega} \hat{\epsilon} 
	(2K+1) j_K(kr) C^{(K)}_{M_K=0}(\hat{r}) \sin(kZ+\tfrac{\pi}{2}K) \;. 
\end{eqnarray}

\end{document}